\documentclass[12pt,preprint]{aastex}

\shorttitle{$G$-band and Hard X-ray Emissions of the 2006 December 14 Flare}
\shortauthors{Watanabe et al.}
\begin{document}

\title{$G$-band and Hard X-ray Emissions of the 2006 December 14 flare observed by {\it Hinode}/SOT and {\it RHESSI}}

\author{Kyoko Watanabe\altaffilmark{1}, S\"{a}m Krucker\altaffilmark{2}, Hugh Hudson\altaffilmark{2}, Toshifumi Shimizu\altaffilmark{1}, \\ Satoshi Masuda\altaffilmark{3} and Kiyoshi Ichimoto\altaffilmark{4}}
\email{watanabe.kyoko@isas.jaxa.jp}

\altaffiltext{1}{Institute of Space and Astronautical Science, Japan Aerospace Exploration Agency, 3-1-1 Yoshinodai, Chuo-ku, Sagamihara, Kanagawa 252-5210, Japan}
\altaffiltext{2}{Space Sciences Laboratory, University of California at Berkeley, 7 Gauss Way, Berkeley CA 94720-7450, USA}
\altaffiltext{3}{Solar-Terrestrial Environment Laboratory, Nagoya University, Furo-cho, Chikusa-ku, Nagoya 464-8601, Japan}
\altaffiltext{4}{Kwasan and Hida Observatories, Kyoto University, Yamashina, Kyoto 607-8471, Japan}

\begin{abstract}
We report on $G$-band emission observed by the Solar Optical Telescope onboard the {\it Hinode} satellite in association with the X1.5-class flare on 2006 December~14.
The $G$-band enhancements originate from the footpoints of flaring coronal magnetic loops, coinciding with non-thermal hard X-ray bremsstrahlung sources observed by the {\it Reuven Ramaty High Energy Solar Spectroscopic Imager}.
At the available 2 minute cadence, the $G$-band and hard X-ray intensities are furthermore well correlated in time. Assuming that the $G$-band enhancements are continuum emission from a blackbody, we derived the total radiative losses of the white-light flare (white-light power).
If the $G$-band enhancements additionally have a contribution from lines, the derived values are overestimates.
We compare the white-light power with the power in hard X-ray producing electrons using the thick target assumption. Independent of the cutoff energy of the accelerated electron spectrum, the white-light power and the power of accelerated electrons are roughly proportional.
Using the observed upper limit of $\sim$30 keV for the cutoff energy, the hard X-ray producing electrons provide at least a factor of 2 more power than needed to produce the white-light emission.
For electrons above $40\,{\rm keV}$, the powers roughly match for all four of the time intervals available during the impulsive phase.
Hence, the flare-accelerated electrons contain enough energy to produce the white-light flare emissions.
The observed correlation in time, space, and power strongly suggests that electron acceleration and white-light production in solar flares are closely related.
However, the results also call attention to the inconsistency in apparent source heights of the hard X-ray (chromosphere) and white-light (upper photosphere) sources.
\end{abstract}


\section{Introduction}

In association with solar flares, we sometimes observe enhancements of visible continuum, in which case the event is termed a ``white-light flare.''
Although white-light events had previously been mainly associated with energetic flares ({\it GOES} X-class), there are now reports of continuum emission from events as weak as C-class flares \citep{Matthews2003, Hudson2006, Wang2009, Jess2008} thanks to accurate photometry from space achieved by {\it Yohkoh}, {\it TRACE}, and {\it Hinode}, and by improved ground-based instruments.
However, white-light flares are still very infrequently observed and some energetic events do not show any enhancement in white light.
The processes causing it remain unclear \citep{Neidig1989}.
Because there is a good correlation of light curves and sites of emission between optical continuum and hard X-rays \citep[e.g.,][]{RustHegwer1975, Neidig1989, Hudson1992, Metcalf2003, Xu2006}, there is some consensus that the origin of white-light emission lies in the energy in accelerated particles, especially non-thermal electrons.

Using the thick-target model \citep{Brown1971}, the energy in flare-accelerated electrons can be compared to the radiative losses in white light \citep[e.g.,][]{Hudson1972}.
If flare-accelerated electrons indeed produce the white-light emission, the energy content in electrons must be larger than the radiative losses in white light.
Due to the (inferred) steep electron spectrum, the energy in electrons strongly depends on the cutoff of the electron spectrum at low energies.
To match the energies, \citet{Neidig1989} and \citet{Ding2003} estimated the cutoff energy of electrons at more than $50\,{\rm keV}$, whereas \citet{Fletcher2007} obtained values below $25\,{\rm keV}$ from a statistical analysis of {\it Transition Region and Coronal Explorer (TRACE)} and {\it Reuven Ramaty High Energy Solar Spectroscopic Imager (RHESSI)} observations.
These differences might be due to the variation from flare to flare.
In any case, a cutoff energy of $\sim20\,{\rm keV}$ can supply the white-light power, but not $100\,{\rm keV}$.  

The Solar Optical Telescope (SOT) of \textit{Hinode} \citep{Tsuneta2008, Suematsu2008, Shimizu2008, Ichimoto2008} makes observations in white light.
Its broadband filter imager (BFI) take images in red ($668.40\,{\rm nm}$, width $0.4\,{\rm nm}$), green ($555.05\,{\rm nm}$, width $0.4\,{\rm nm}$) and blue ($450.45\,{\rm nm}$, width $0.4\,{\rm nm}$) continuum ranges.
Radiation at these wavelengths comes from the photosphere and hence reflects the broadband continuum emission well.
However, SOT normally obtains only infrequent images in these filters.
More frequently, SOT takes images in the $G$-band ($430.50\,{\rm nm}$, width $0.83\,{\rm nm}$), formed mainly from CH~line opacity.
\cite{Carlsson2007} show contribution functions for these filters; the $G$-band has a photospheric and an upper-photospheric contribution.
It therefore serves well to define the morphology of white-light flares and it was also used in the \textit{Yohkoh} observations \citep{Hudson1992,Matthews2003}.
However, $G$-band emission could contain not only continuum emission, but also CH~line emission.
If the $G$-band emission contains line emission, the radiative losses estimated from the $G$-band emission assuming blackbody radiation is overestimates of the true losses.
However, in this paper, we treat the $G$-band emission mainly came from the continuum emission, and we therefore use $G$-band images as a proxy for the white-light images.

SOT observed white-light emission from three X-class flares in 2006 December \citep{Wang2009}.
$G$-band emission of the largest event (X3.4 flare on 2006 December~13) is reported by \citet{Isobe2007} and \citet{Jing2008}.
\citet{Isobe2007} concluded that the white-light emission could be produced by radiative back-warming resulting from particle-beam heating, and \citet{Jing2008} noted that the white-light emissions appeared at the sites of the largest inferred reconnection rates.
In this paper, we describe the white-light observations of the 2006 December~14 flare that was also observed by the {\it RHESSI} \citep{Lin2002}.
We obtain X-ray energy spectra for each foot-point separately, using {\it RHESSI} imaging spectroscopy, and compare the results with energy estimated from {\it Hinode} $G$-band images (Section \ref{observation}).
In Section \ref{relation}, energy estimates are discussed for different cutoff energies.

\section{Observations by {\it Hinode}/SOT and {\it RHESSI}}
\label{observation}

\begin{figure}[tbp]
\epsscale{0.8}
\hspace*{-5mm} \plotone{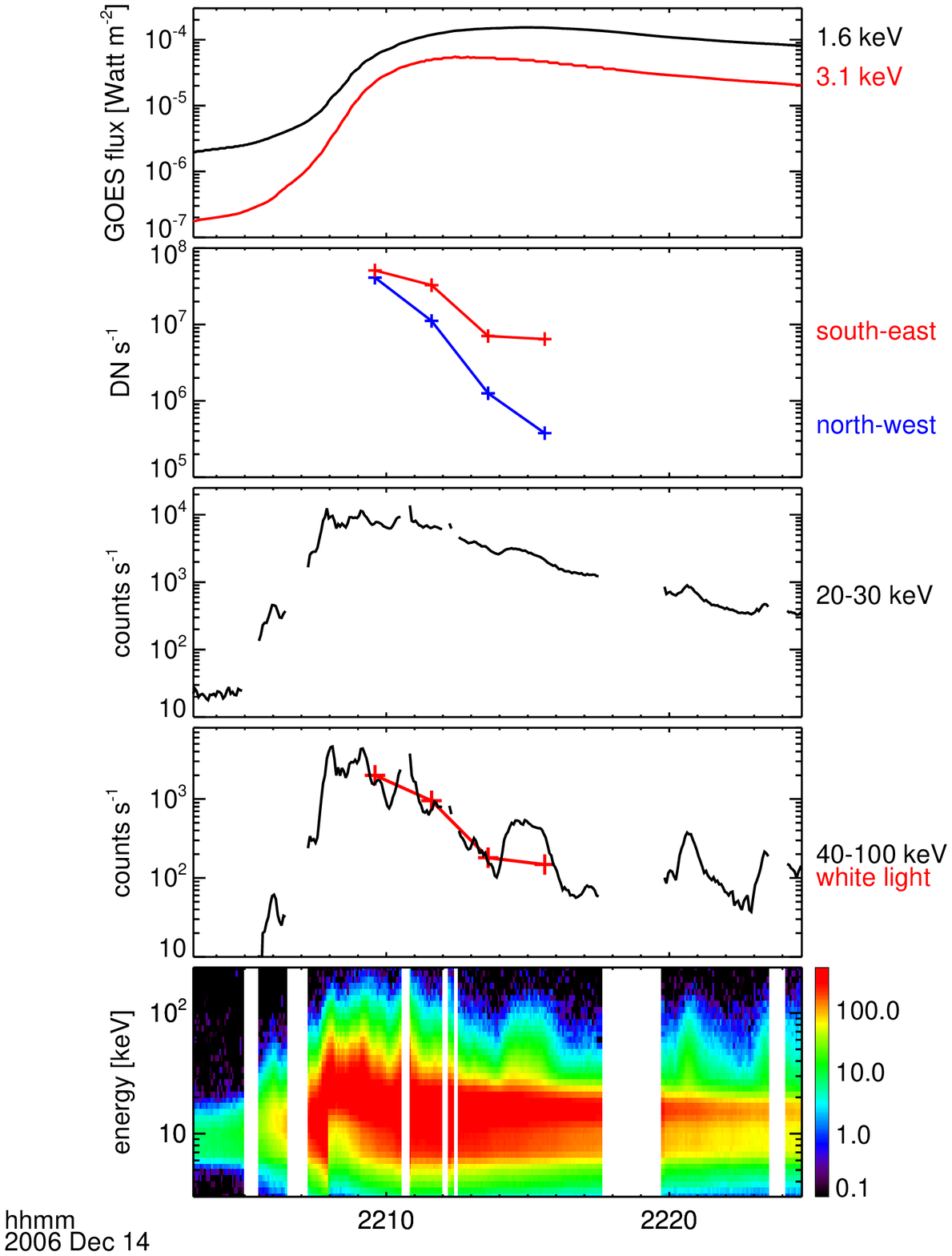}
\caption{Light curves of {\it GOES} soft X-rays, $G$-band excess emission from {\it Hinode}/SOT, and $20 - 30$, $40 - 100\,{\rm keV}$ range hard X-ray emissions from {\it RHESSI}.
The bottom panel is a {\it RHESSI} spectrogram, and the color bar on the right shows the spectral flux (counts/4sec).
Total white-light emission (sum of south-east and north-west emission at each image time) is overlaid on hard X-ray light curve in the next-to-bottom panel, showing a good correlation.}
\label{rhessi_lc}
\end{figure}

On 2006 December~14, an X1.5-class flare occurred in active region NOAA~10930; this region produced a total of three X-class flares in 2006 December.
{\it Hinode}/SOT took $G$-band images every 2 minutes, obtaining four images between 22:09 and 22:16\,UT that showed flare-related brightenings.
The exposure times of these images were 0.031~s, and the spatial binning was 0.109~arcsec/pixel.
To estimate the excess emission from these images, we made a reference image by averaging images before (22:07\,UT) and after (22:17\,UT), and subtracted the reference from the flare images.
We then calculated background pixel statistics for these difference images, and estimated the excess emission by summing the pixels at more than $3 \sigma$ (= 245.3~DN/exposure time/pixel) above the background level.
{\it RHESSI} had full coverage of this flare.
Figure~\ref{rhessi_lc} shows light curves from {\it GOES} X-rays, $G$-band emission by {\it Hinode}/SOT obtained as described above, and hard X-rays observed by {\it RHESSI}.
The total white-light emission correlates well with hard X-rays at $40-100\,{\rm keV}$, as shown in the fourth panel of Figure~\ref{rhessi_lc}.
Note that the white-light images are brief samples, whereas the hard X-ray record is continuous.

\begin{figure}[tbp]
\vspace*{-1.5cm}
\epsscale{0.80}
\plotone{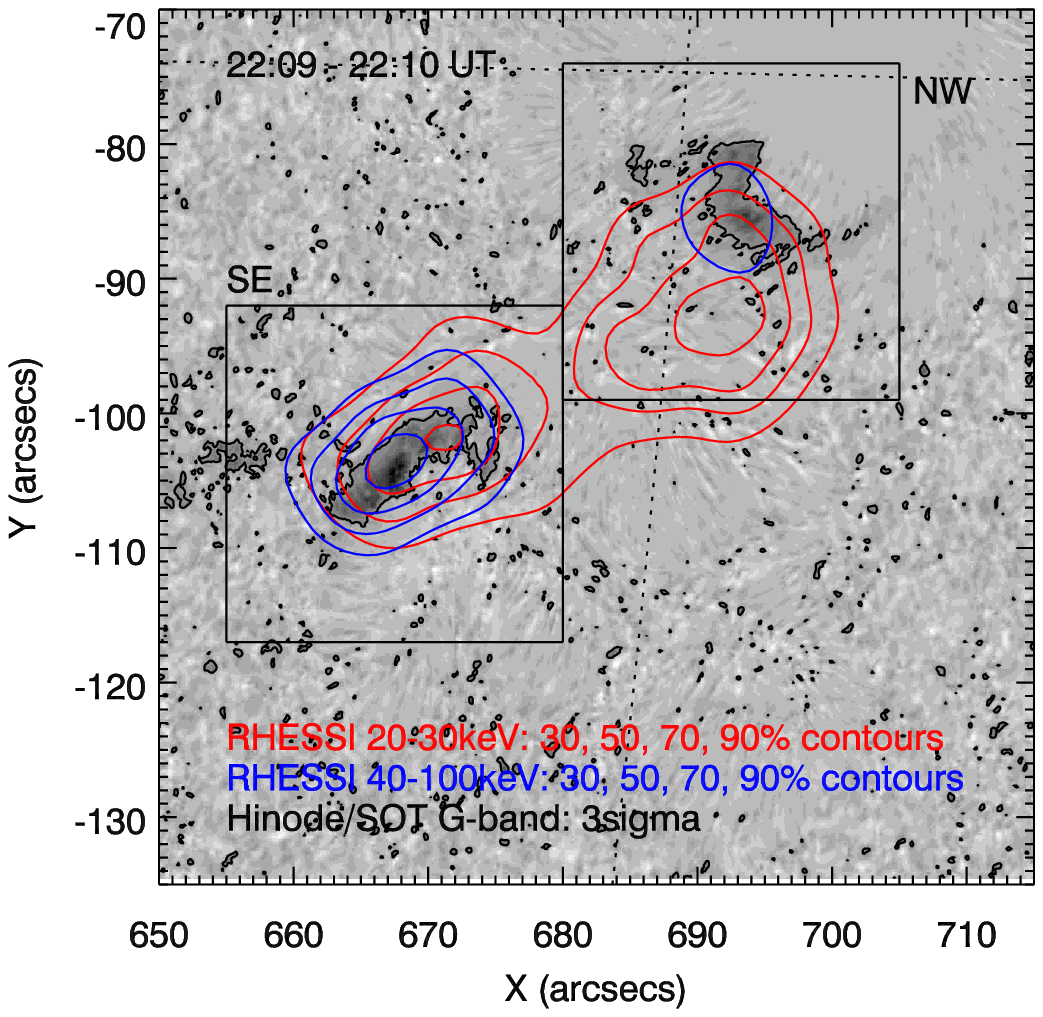}
\plotone{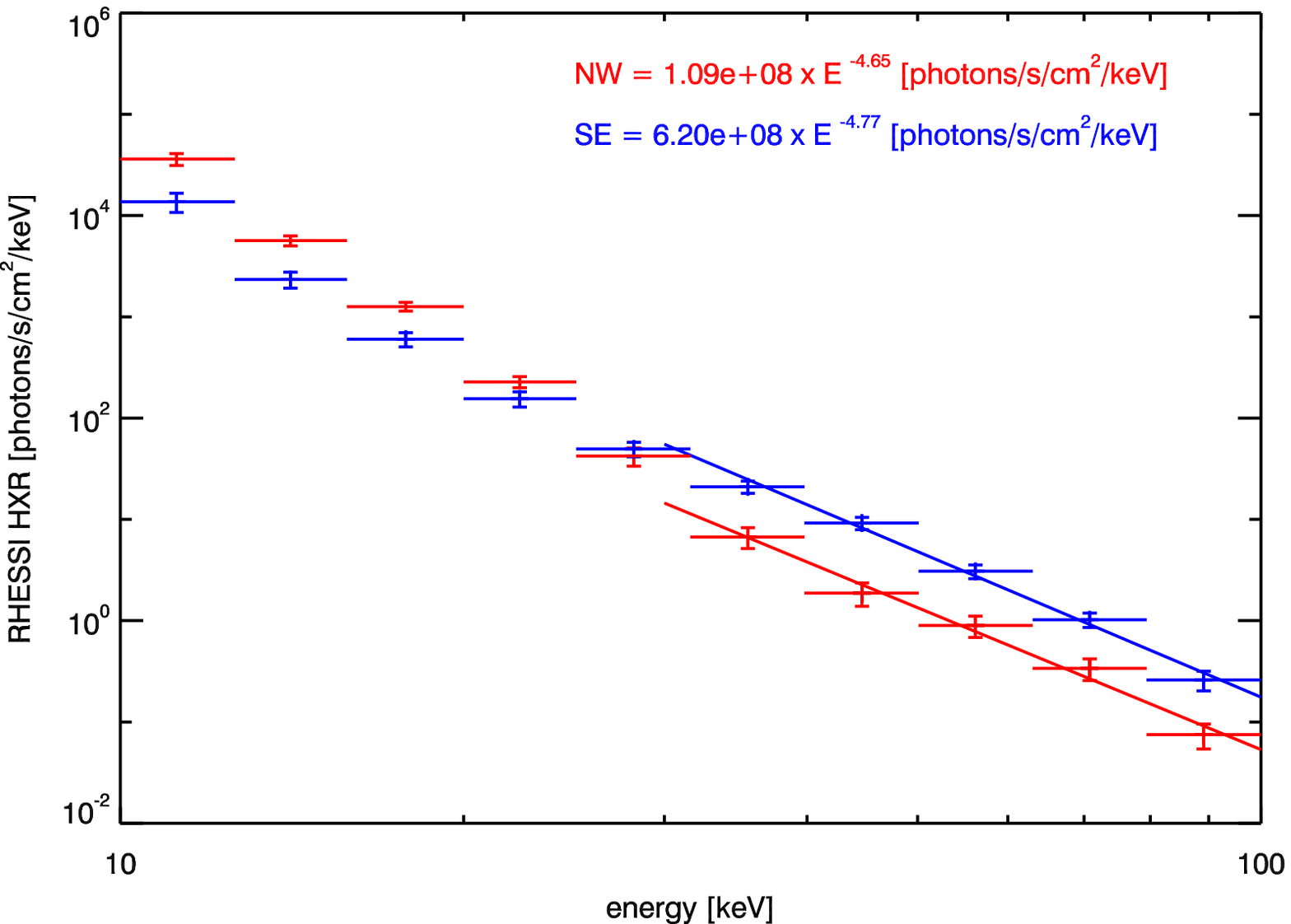}
\caption{{\it Top}: Image of $G$-band emission taken by {\it Hinode}/SOT and {\it RHESSI} hard X-ray contours at 22:09UT.
The background image is the differential $G$-band image (the average of the images taken at 22:07\,UT and 22:17\,UT is subtracted), and the black contours indicate $3 \sigma$ above background. Red contours show $20-30\,{\rm keV}$ emission, and blue contours show $40-100\,{\rm keV}$ emission. The black box give the area over that the {\it RHESSI} image is searched for emission from the footpoint. Only emission above $3 \sigma$ is added up to derive the spectrum.
{\it Bottom}: Photon spectra for each foot-point at 22:09UT.
The red crosses and lines indicate spectra in the NW~box, and blue ones indicate spectra in the SE~box, with $1\sigma$ error bars. Energy bands of spectrum were in a logarithmic space and each spectrum was fitted in the range above $30\,{\rm keV}$.}
\label{gb_rhessi_1}
\end{figure}

\begin{figure}[tbp]
\epsscale{0.80}
\plotone{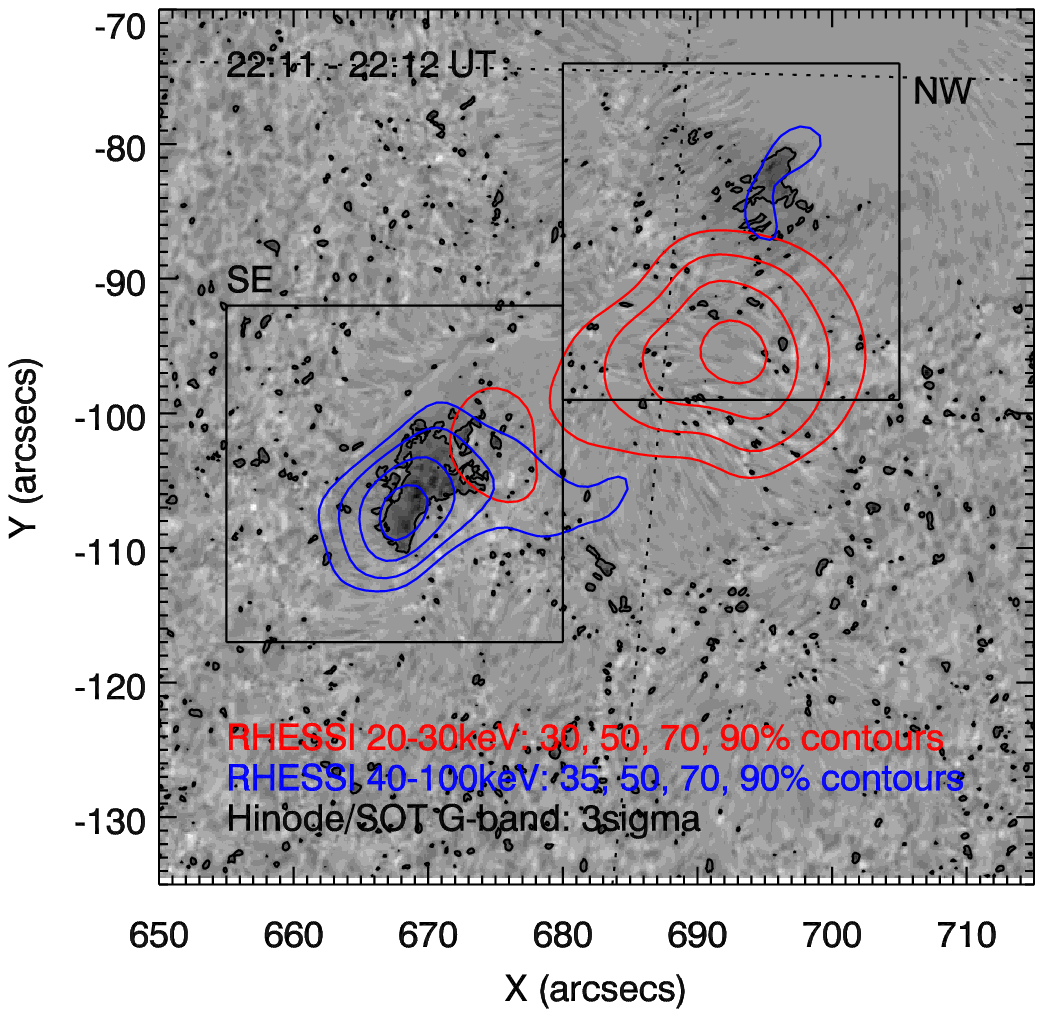}
\plotone{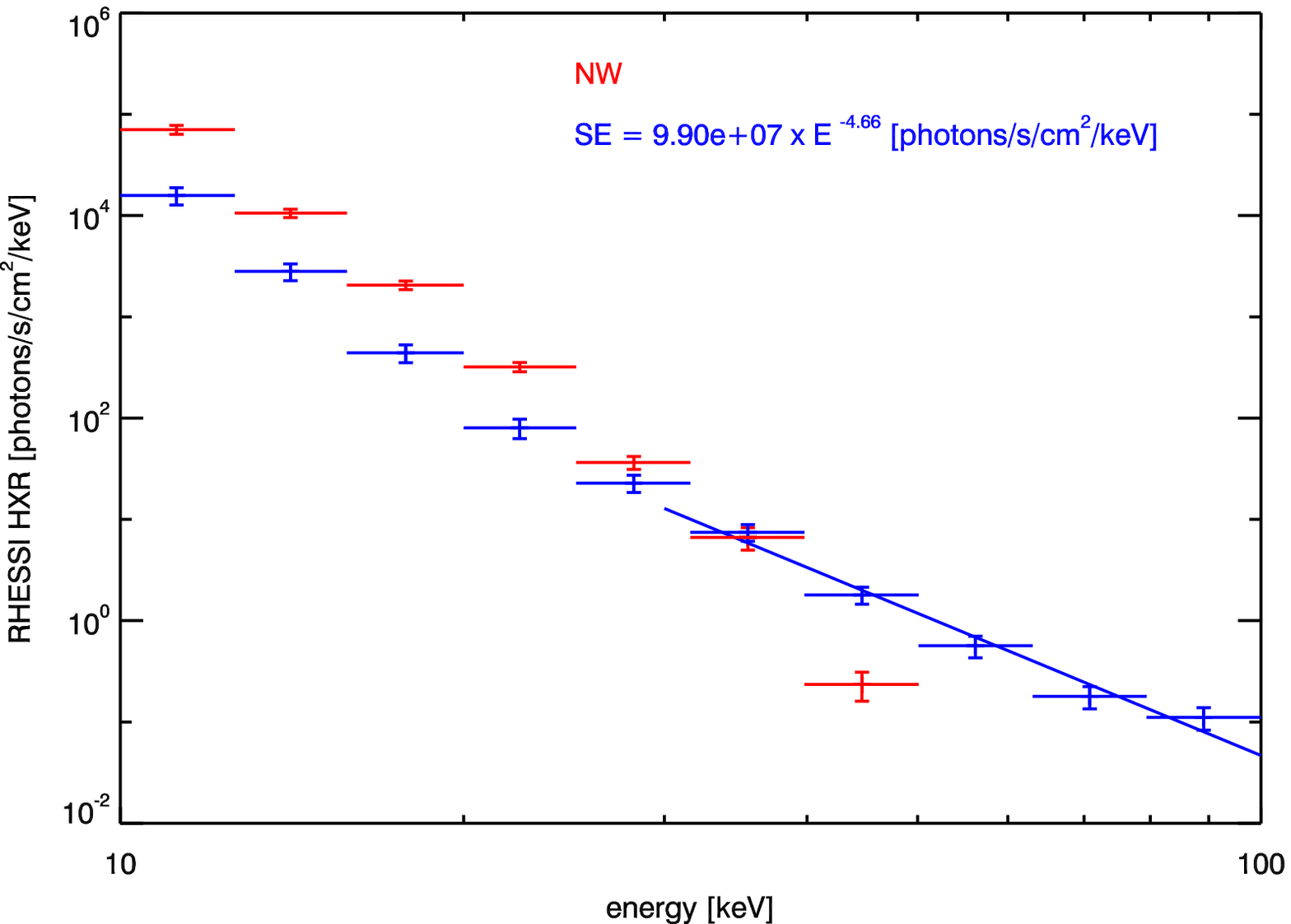}
\caption{Same as in Figure~\ref{gb_rhessi_1} for 22:11UT.}
\label{gb_rhessi_2}
\end{figure}

\begin{figure}[tbp]
\epsscale{0.80}
\plotone{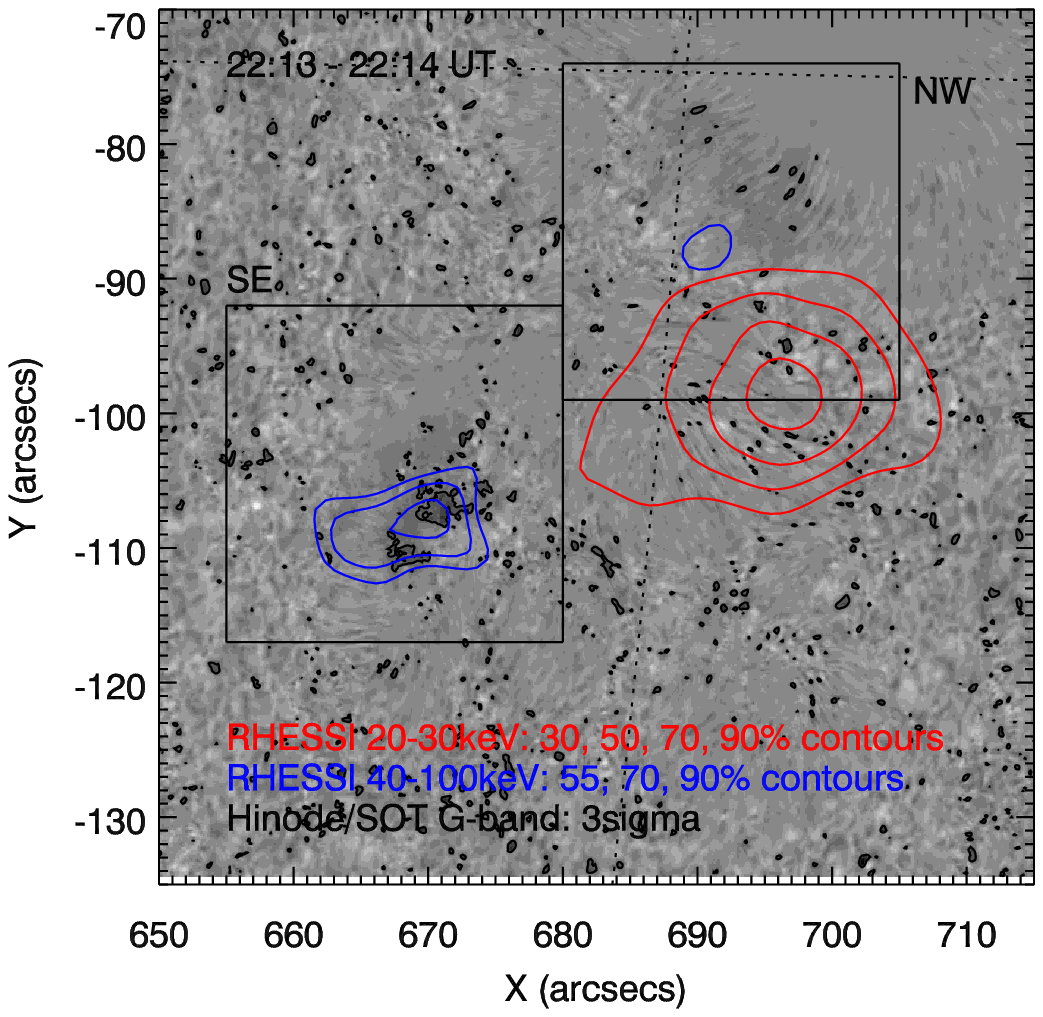}
\plotone{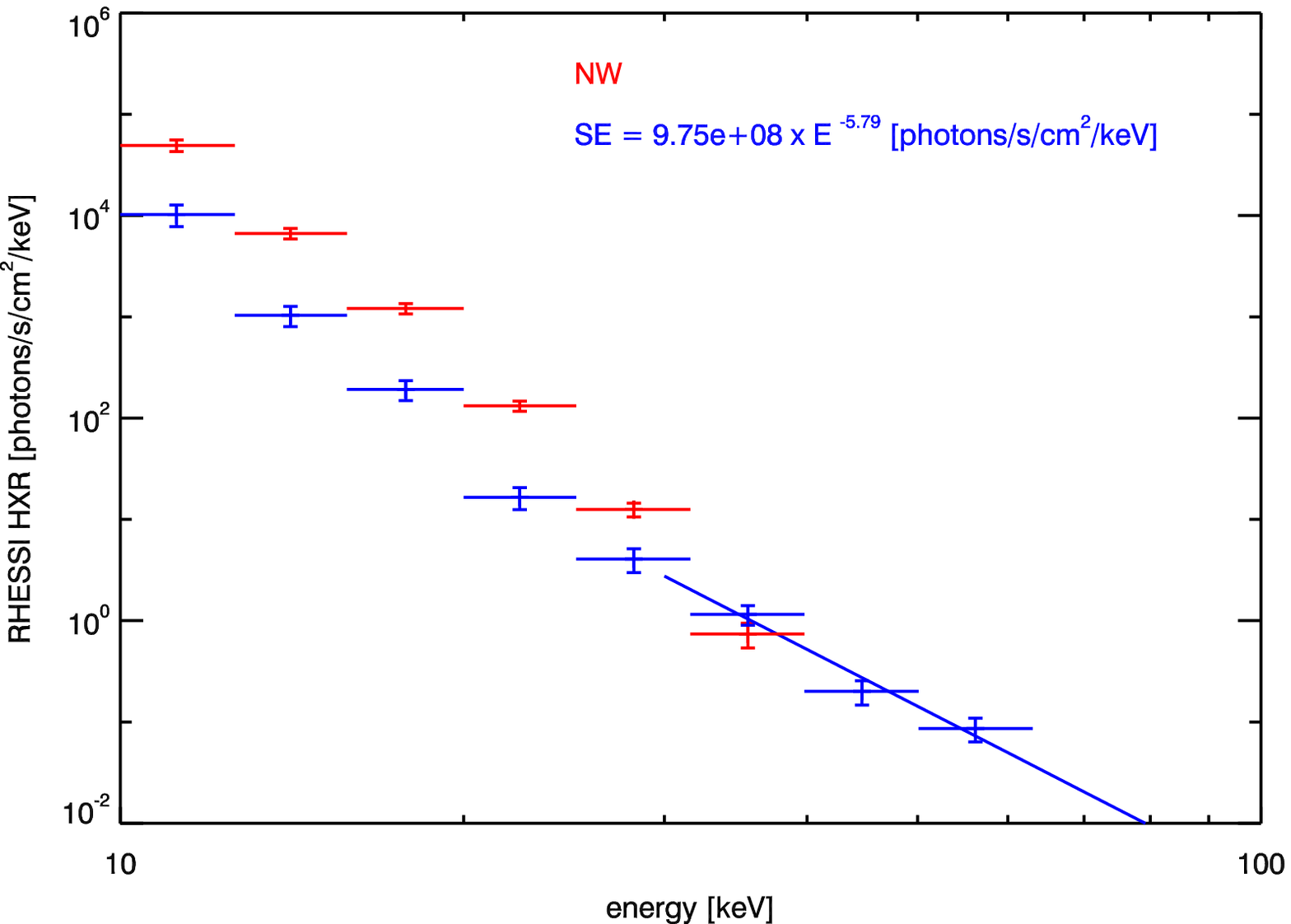}
\caption{Same as in Figure~\ref{gb_rhessi_1} for 22:13UT.}
\label{gb_rhessi_3}
\end{figure}

\begin{figure}[tbp]
\epsscale{0.80}
\plotone{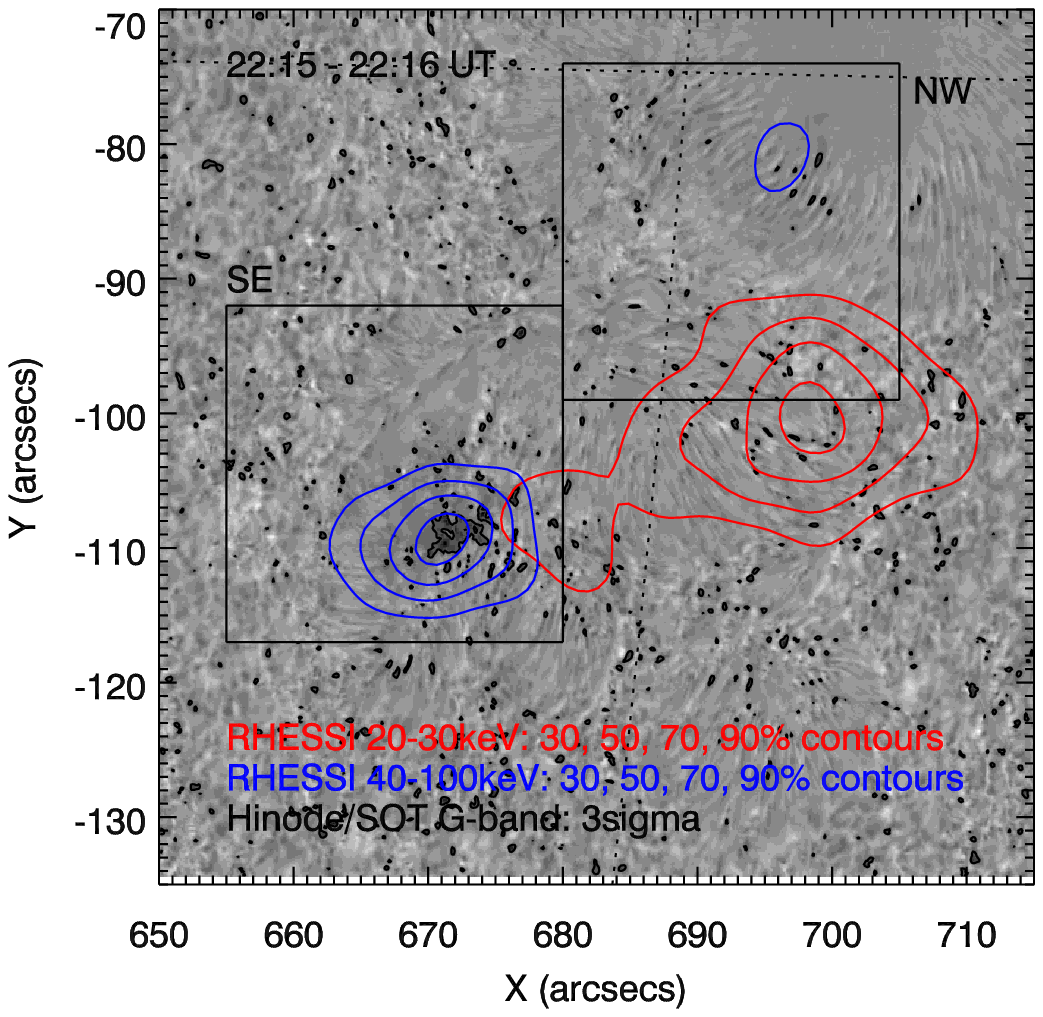}
\plotone{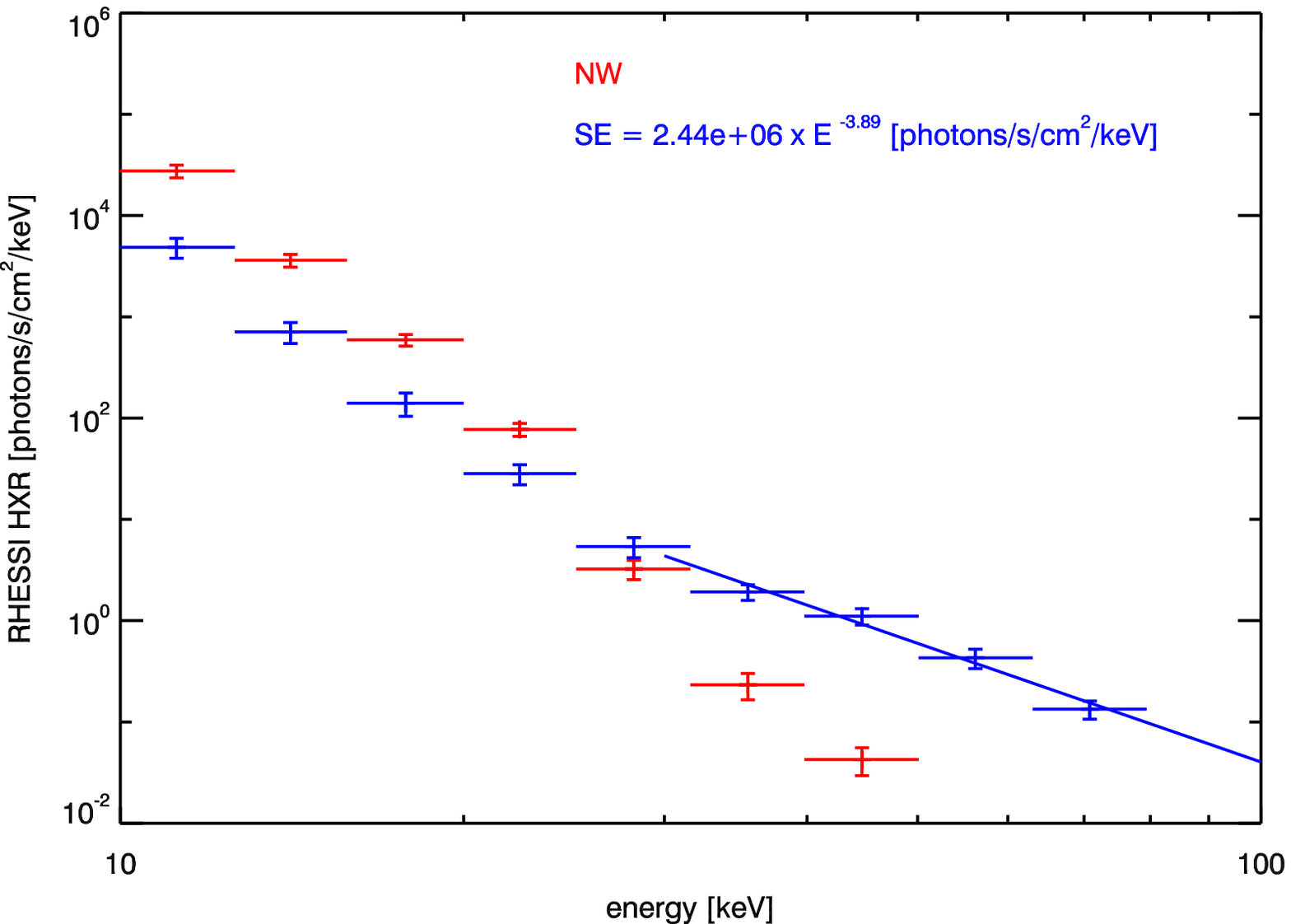}
\caption{Same as in Figure~\ref{gb_rhessi_1} for 22:15UT.}
\label{gb_rhessi_4}
\end{figure}

Around the times of the {\it Hinode} images, we made {\it RHESSI} hard X-ray images integrated over 1 minute by using the CLEAN algorithm with sub-collimators 2, 3, 4, 6, 8, and~9.
Despite its lower energy resolution, sub-collimator 2 is used here to obtain higher spatial resolution for images reconstructed over a broad energy range.
Sub-collimator 5 is affected by cumulative radiation damage and is therefore not used (the observations presented here were taken before the first {\it RHESSI} anneal).
It is necessary to register the {\it RHESSI} and {\it Hinode} images empirically.
For this purpose we compared {\it RHESSI} thermal X-ray images at 6 keV with images from {\it Hinode}/XRT and shifted the latter to fit.
The alignment between SOT/$G$-band and XRT was corrected using the method of \citet{Shimizu2007}.
The results for this registration are shown in the top figures of Figure~\ref{gb_rhessi_1}-\ref{gb_rhessi_4}.
Above 40 keV, the hard X-ray images show two foot-point sources, with the $G$-band emission at the same locations within uncertainties.
The hard X-ray footpoint regions have different spectral behavior; the $20-30\,{\rm keV}$ energy range shows the northwest (NW) source more clearly than the southeast (SE) one.
The behavior in white light is opposite to this, i.e., the SE region is brighter.
On the other hand, if we use $40-100\,{\rm keV}$ hard X-rays, a similar feature of the white-light emission was obtained.
This SE feature was seen clearly in all four images at $40-100\,{\rm keV}$.

We derived energy spectra of each footpoint separately by obtaining images at different energy bands \citep[e.g.,][]{KruckerLin2002}.
For the first time interval, the spectrum for both footpoints could be derived showing similar power-law spectra (Figure~\ref{gb_rhessi_1}).
For the later time intervals, the NW footpoint is too weak to accurately determine its spectrum, and only the spectrum of the SE footpoint is derived (Figures~\ref{gb_rhessi_1}-\ref{gb_rhessi_4}).

\section{Relationship between white-light and hard X-ray emissions}
\label{relation}

In this section, the power of the white-light continuum (radiative losses) is compared with the energy deposition by flare-accelerated electrons producing the hard X-ray emission.
The white-light power is calculated assuming that the $G$-band enhancements are continuum emission from a blackbody.
This is a strong assumption that is not necessarily fulfilled.
Besides the continuum emission, the $G$-band emission could contain line emission.
In this case, our assumption leads to an overestimation of the white-light power.
A further limitation of our assumption is that the spectrum of the white-light flare emission could be different from a blackbody.
We further assume temperatures of $6000\,{\rm K}$ for the quiet Sun, $5500\,{\rm K}$ for the penumbral regions, and $4200\,{\rm K}$ for the umbra \citep[e.g.,][]{Cox2000}, and calculate the temperature of the white-light flare source using the response of the $G$-band filter and Planck's law ($B_{\lambda}(T) = 2hc^2/\lambda^5 (e^{hc/\lambda kT}-1)$).
The power of the white-light continuum $E$ (i.e., radiative losses) from the temperature $T_e$ is then calculated using Stefan-Boltzmann's law ($E=5.67 \times 10^{-8}\,T_e^4$).
Note that this represents a lower limit because the continuum could extend into the UV, e.g. via the Balmer continuum \citep[e.g.][]{Fletcher2007}.

\begin{figure}[tbp]
\epsscale{0.50}
\plotone{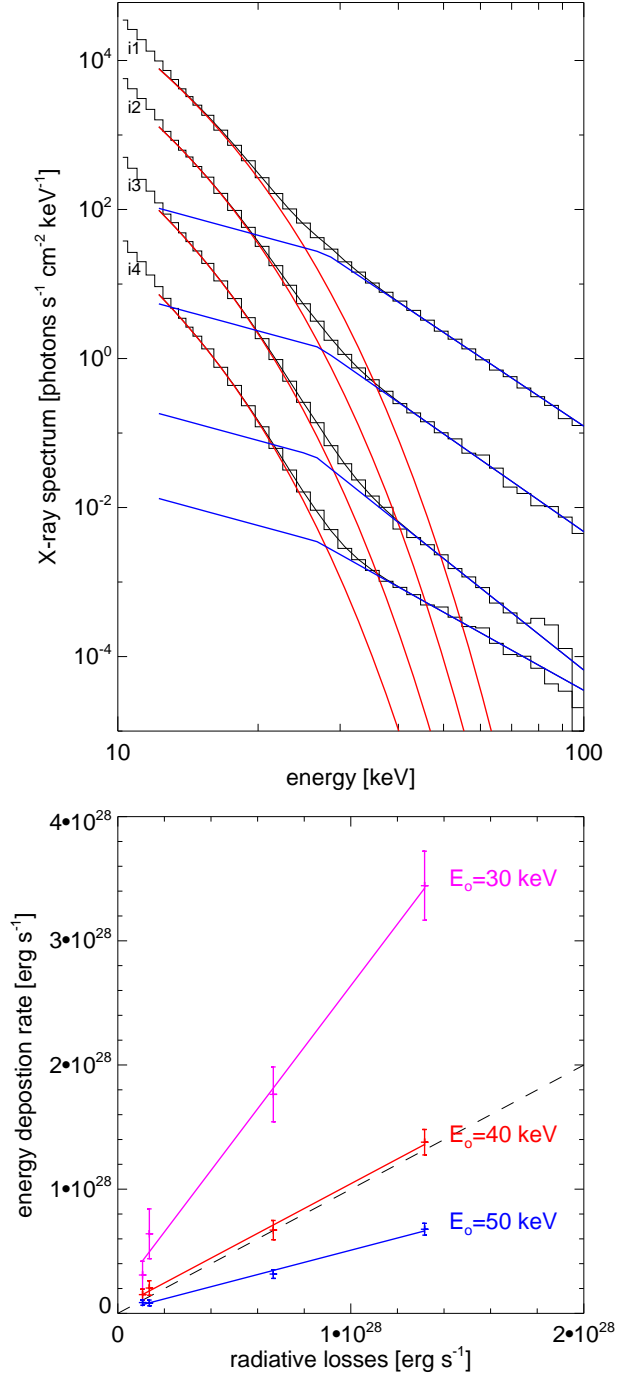}
\caption{{\it Top}: Hard X-ray photon spectra integrated over 4 seconds around the time of the four $G$-band images.
The black histogram represents the data.
The red and blue curves are a thermal and broken power law fit to the data, while the black curve represents the sum of the two fits. Spectra from later intervals are successively divided by an additional factor of 10 for a clearer representation.
{\it Bottom}: Correlation plot of total power in white-light continuum (radiative losses) and the derived power in non-thermal electrons for different cutoff energies as indicated.
The four data points correspond to the four time intervals and are connected by lines to indicate time evolution.
The dashed line outlines equality between energy deposition rate and radiative losses.}
\label{RHESSI_4sec}
\end{figure}

\begin{figure}[tbp]
\epsscale{1.00}
\plotone{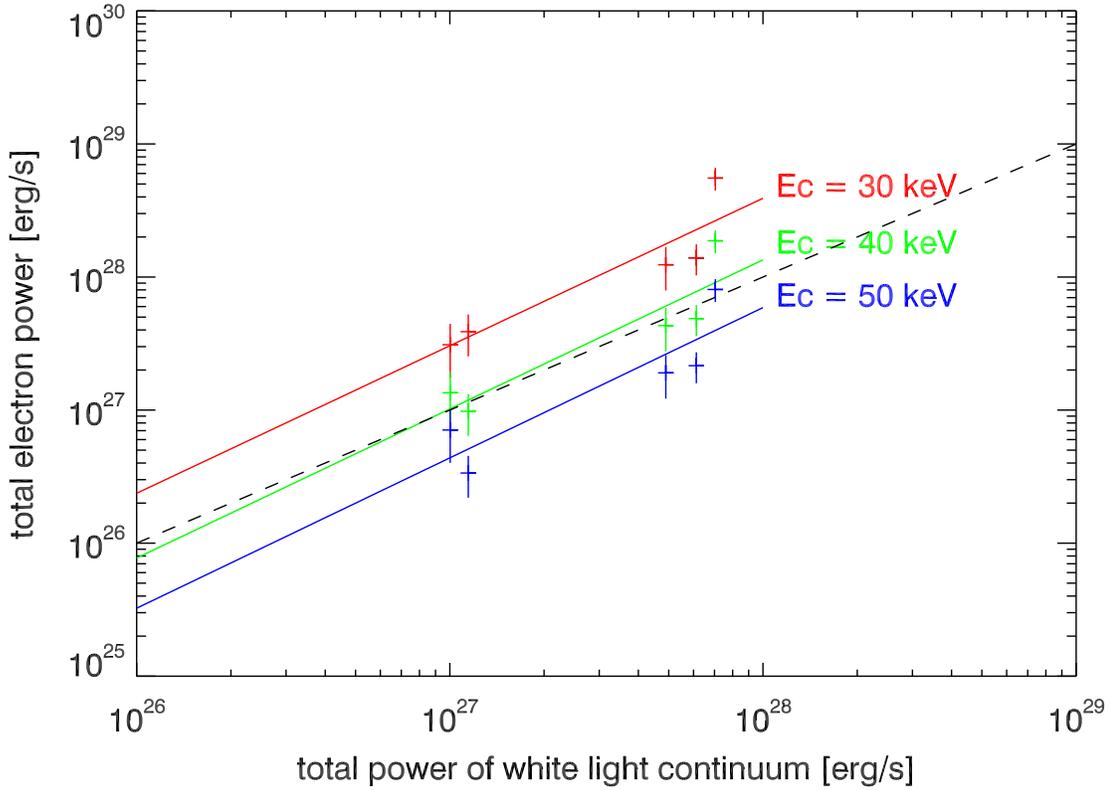}
\caption{Same plot as Figure~\ref{RHESSI_4sec} (bottom) derived from imaging spectroscopy results.
The shown data points correspond to the NW footpoint (first interval only; dynamic range of hard X-ray images is too low to derive accurate values for later intervals) and the SE footpoint (all intervals) as shown in the images of Figure~\ref{gb_rhessi_1} to \ref{gb_rhessi_4}.
The different colors correspond to different cutoff energies as indicated.
}
\label{cutoff_ene}
\end{figure}

The total power in non thermal electrons $P$ above a given cutoff energy $\varepsilon_c$ in the thick-target approximation \citep[Equation~(\ref{eq1});][]{Hudson1978} can be derived from the observed hard X-ray photon spectrum $I(\epsilon_x)$ (as shown in Figures~\ref{gb_rhessi_1}-\ref{gb_rhessi_4}):
\begin{eqnarray}
P(\varepsilon \ge \varepsilon_c) = 4.3 \times 10^{24}\,\frac{b(\gamma)}{\gamma-1}\,A\,\varepsilon_c^{-(\gamma-1)} \hspace{1cm} {\rm (erg/s)}, \label{eq1} \\
I(\epsilon_x) = A\,\epsilon_x^{-\gamma} \hspace{1cm} {\rm (photons/cm^2/s/keV)}. \label{eq2}
\end{eqnarray}
The factor $b(\gamma)$ is an auxiliary function from \citet{Brown1971} as calculated by \citet{Hudson1978} for a relevant range of spectral indices $\gamma$ as $b(\gamma) \approx 0.27 \gamma^{3}$.
Two approaches to derive $P$ are presented:
first, $P$ is estimated from the total flare spectrum integrated over 4 seconds around the time of the $G$-band image (Figure~\ref{RHESSI_4sec}).
Second, $P$ is estimated for each footpoint source separately using imaging spectroscopy (Figures~\ref{gb_rhessi_1}-\ref{gb_rhessi_4} and \ref{cutoff_ene}).
The advantage of the first approach is that it provides smaller uncertainties and that the {\it RHESSI} images are taken almost simultaneously with the $G$-band images (cf. Figure~\ref{rhessi_lc}).
The second approach compares individual sources, but a time integration of 60 seconds is needed to obtain a significant result.
However, both approaches give similar results (Figures~\ref{RHESSI_4sec} and \ref{cutoff_ene}).  
Error bars are estimated from the uncertainties of the fit parameters. The uncertainty in $\delta$ (typically $\sim$5\% for the spatially integrated spectra, and 10\% for imaging spectroscopy \citep{KruckerLin2002}) dominates the error budget.
There is a clear correlation between the power in white-light emissions and the energy deposition rate by non-thermal electrons for all cutoff energies.
For low cutoff energies, the power provided by non-thermal electrons is well above the white-light power (e.g. $\sim$50\% of the power in electrons above $30\,{\rm keV}$ is enough to account for the white-light emission).
The total power of electrons above $\sim 40\,{\rm keV}$ roughly matches the white-light power.
If the $G$-band emission contains not only continuum emission but also line emission, the values for the white-light power derived above are overestimates.
In any case, for all realistic cutoff energies, the energy in non thermal electrons (derived from thick-target assumptions) is larger than the white-light power, and independent of values of the cutoff energies, the derived powers are correlated.

\section{Discussion and Summary}

The observations discussed in this paper show that the solar flare white-light emission is closely related in time, space, and power to the acceleration of non-thermal electrons.
To explain the observed correlation between white light and high energy hard X-ray emission in the simplest possible way, the two components should originate in the same source region.
Continuum emission in the $G$-band emission comes from $0-100\,{\rm km}$ above the photosphere \citep[see Figure~1 of ][]{Carlsson2007}, and hard X-ray emission in $50-100\,{\rm keV}$ originates in the chromosphere.
Observationally, the emission site of $50-100\,{\rm keV}$ hard X-rays is estimated at $6.5 \times 10^{3}\,{\rm km}$ height above the photosphere from (early) {\it Yohkoh} observations \citep{Matsushita1992} and around $600\,{\rm km}$ height by  \textit{RHESSI} for a single event \citep{Kontar2008}.
This information is weak and we would like to see systematic \textit{RHESSI} data analyses on this point, but the existing data suggest a difference of more than $500\,{\rm km}$ between the emission sites \citep{Kontar2008, Carlsson2007}.
Theoretically, a $50-100\,{\rm keV}$ electron should thermalize some $1000\,{\rm km}$ height above the photosphere; at this mid-chromospheric height, the density is about $10^{13.5}/{\rm cm}^3$ \citep{Neidig1989}.
At these energies, however, the electrons cannot penetrate into the lower chromosphere, and thus they do not heat the photosphere.
Electron energies more than $900\,{\rm keV}$ are necessary for penetration to the photosphere, even if the flare site has become ionized \citep{Neidig1989}.
However, the energy in $900\,{\rm keV}$ electrons is far too small (by about 4 orders of magnitude, assuming the power law seen at $40\,{\rm keV}$ can be extrapolated to $900\,{\rm keV}$) to produce the white-light emission.
The data presented here therefore call attention to the need for a white-light emission model which can explain the good correlation with high-energy electron emission and difference of the emission height of white light and hard X-rays.
Non thermal ionization levels enhance the continuum \citep{Hudson1972} and also make back-warming possible \citep[e.g.,][]{Metcalf2003}, but we do not have well-defined models for these processes in realistic physical conditions yet.



\acknowledgments
{\it Hinode} is a Japanese mission developed and launched by ISAS/JAXA, collaborating with NAOJ as a domestic partner, NASA and STFC (UK) as international partners.
Scientific operation of the {\it Hinode} mission is conducted by the {\it Hinode} science team organized at ISAS/JAXA.
This team mainly consists of scientists from institutes in the partner countries.
Support for the post-launch operation is provided by JAXA and NAOJ (Japan), STFC (U.K.), NASA, ESA, and NSC (Norway).
{\it RHESSI} is the NASA Small Explorer mission, we appreciate the {\it RHESSI} team, for their support to the mission and guidance in the analysis of the {\it RHESSI} satellite data. 
Authors Hudson and Krucker thank NASA for support under grant NAS 5-98033.
A part of this work was carried out by the joint research program of the Solar-Terrestrial Environment Laboratory, Nagoya University.
We also appreciate fruitful comments from an anonymous referee.

\end{document}